# Numerical Method for Free Electron Laser using an Overmoded Rectangular Waveguide

Weiwei Li, Heting Li, Zhigang He[*], Qika Jia, Lin Wang and Yalin Lu[*]

*National Synchrotron Radiation Laboratory, University of Science and Technology of China, Hefei, Anhui, 230029, China*

Numerical simulation codes are basic tools for designing Free Electron Lasers (FELs). This paper describes a numerical method for the time-dependent, three-dimensional simulation of the free electron laser (FEL) using a rectangular waveguide within overmoded configuration when the radiation wavelength is much shorter than the waveguide cut-off wavelength. Instead of developing a new code, the GENESIS simulation code is modified for our purpose. This method presented here can be used for extending the capacity of GENESIS to cover this special configuration. The major modification is to apply the metal boundary conditions on the field equations in a limited rectangular region and the full Cartesian mesh using the Alternating Direction Implicit (ADI) integration scheme to solve the field equation remains adopted.

## I. Introduction

The far-infrared (FIR) and terahertz FELs, which usually use a undulator as the radiation medium, have long been in focus due to their many important applications and the small number of sources of coherent radiation available in the spectral region [1]. For practical purposes, in most cases the undulator is located outside the vacuum chamber. Therefore, the size of the vacuum chamber (or the space between the undulator poles if the chamber is outside) is limited by the need of producing a sufficient magnetic field within the undulator. However, due to the diffraction effect, the transverse size of the optical mode tends to be lager and the vacuum chamber is usually performed as an optical waveguide at the relatively long wavelengths.

The waveguide effect on FEL has been studied by a number of authors [2-13]. In a waveguide FEL, the radiation field propagates inside the waveguide as waveguide modes, not like a light flux in a free space FEL. This characteristic behavior of the radiation field in a waveguide FEL makes intuitive understanding and complete simulation of the waveguide FEL very difficult. For a general numerical method, one should use eigenmode decomposition for studying the detuning parameter with the radiation frequency, the slippage distance between the electron pulse and the radiation pulse, and the interaction of electrons with the radiation field for each mode, et al.

In this paper, we present a numerical method for the time-dependent, three-dimensional simulation of the free electron laser (FEL) using a rectangular waveguide within overmoded configuration. Instead of developing a new code, the GENESIS simulation code is modified for our purpose. When an overmoded waveguide is used, the radiation wavenumber and the slippage distance are nearly consistent with that of the situation of free space [9]. Thus the most noticeable difference between the waveguide FEL and its free-space counterpart will be the transverse field evolution dependence on the waveguide and the calculation of the amplitude and phase for each mode becomes unnecessary.

For simplification, we only consider the case of the FEL using the planar undulator and an overmoded rectangular waveguide and this condition can be satisfied for many practical far-infrared (FIR) and terahertz FEL facilities. As an example, for the "CLIO" infrared FEL [14], the transverse size of the waveguide is 18×35 mm, and the relevant cut-off wavelength is $\lambda_c$ = 2.6 cm for $TEM_{11}$ mode which is far from the laser spectral range: 5 um–150 um. As a consequence, the laser mode inside the waveguide can be regarded as widely overmoded.

## II. Numerical method

GENESIS [15] is a time-dependent three-dimensional FEL code and widely used for FEL simulation. The code allows us to study many important effects include details of the electron longitudinal and transverse particle distributions, external focusing for long undulators, effects of errors in the magnetic fields or other system components, longitudinal variation of the undulator field to enhance radiation power, diffraction effects, startup from spontaneous radiation and so on. Thus, instead of developing a new code, the GENESIS simulation code is modified for our purpose.

In the GENESIS simulation, a Cartesian coordinate system is used, where the z-axis coincides with the undulator axis. The transverse coordinates x and y are chosen so that the magnetic field for a planar undulator is parallel with the y-axis. The paraxial approximation of Maxwell equation for the steady-state radiation field can be written as [16]

$$\left[\partial_x^2 u + \partial_y^2 u + 2ik\frac{\partial}{\partial z}\right]u = S \qquad (1)$$

where $u$ is the complex representation of the radiation field, $k$ is the wavenumber, and $S$ is source term (emission from the electron beam). The radiation field is defined at every point of a finite square transverse plane. It uses a discretization of the field on a Cartesian grid ('finite differences') with uniform spacing $\Delta$. The boundary conditions can be chosen to be either Dirichlet or Neumann boundary condition. For the free space propagation, the grid should be lager enough so that no significant amplitude of the radiation field will reach the boundary of the grid and be reflected backwards.

When an overmode rectangular waveguide is used, the simulation region and boundary conditions should be modified accordingly as shown in Fig.1. Note that only the vertical size of the vacuum chamber is limited by the planar undulator gap, the horizontal vacuum chamber size is usually larger. Therefore, we only consider this situation. The square

---

region surroundings by the black solid lines is the situation of the origin GENESIS code and the rectangular region surroundings by dotted lines is the waveguide region. There are $N_C$ and $N_y$ discrete grid points along x and y axis inside the waveguide respectively and the spacing of the grid points remains uniform, and the size of the waveguide is then decided by $N_C$, $N_y$ and the grid resolution $\Delta$. Because many other items, like the electron transverse position, the field diagnostics, and the format of the radiation field, are all related to the definition of the grid, we still keep $N_C$ discrete grid points along y axis in order to be consistent with GENESIS, and the field outside the waveguide is set to be zero.

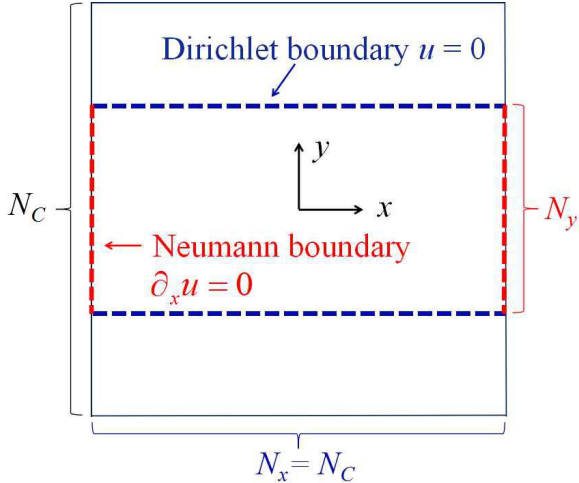

Fig.1 The modified simulation region and boundary conditions

Major modifications are made to solve the steady-state radiation field with the metal boundary conditions as discussed above. Following the GENESIS simulation code, the ADI integration scheme [17, 18] remains adopted, where for each dimension a fully implicit substep is done, while the rest of the dimensions are treated explicitly. This reduces the system into a set of multiple field equations with tridiagonal shape of the Laplace operator. As for the perfect conductor, the magnetic field can be calculated at any positions z using the numerical method above. Then the ohmic losses can be easily calculated using the well known good conductor approximation [19].

### III. Simulation Example

In terms of the electromagnetic field propagation problems (without source term in Eq. 1), Prazeres [20] developed a numerical method using the fast Fourier transform, similar with the eigenmode decomposition. To make a comparison, we use their parameters and give a similar example of propagation of a Gaussian wave through the waveguide, for various distances L as shown in Fig.2. Comparison of the results with those shown in Fig.12 of Ref. [20] shows excellent agreement. Besides the effects of the electron beam can be easily taken into consideration using our approach.

To illustrate some of the capabilities of the proposed approach, We perform a time-dependent simulation of a prebunched terahertz - FEL. An electron bunch train is initiated on the photocathode of a RF gun by a laser pulse train with a time spacing of 1 ps and subsequently accelerated in the rf gun and the linear accelerator (linac). Narrow-band THz radiation is excited as the electron bunch train traveling along the undulator. A detailed information of the electron beam injector can be found in [21]. The total beam change is 240 pC. The cross section and the energy spread distributions the undulator entrance are shown in Fig.3.

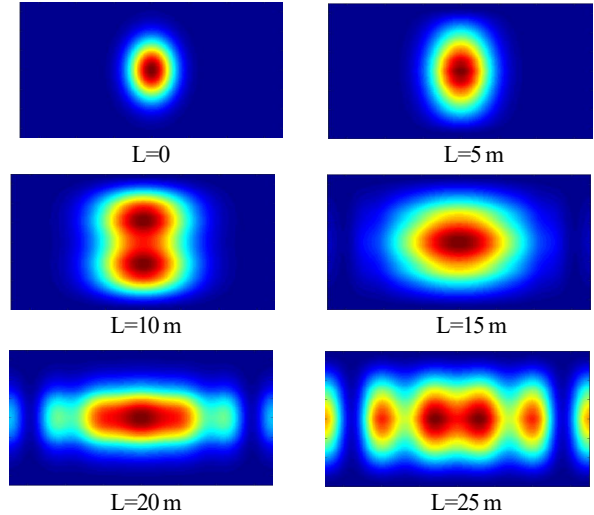

Fig.2 The modulus electric field $E_X(x, y)$ profile in the waveguide area {18× 35} mm at various distances L from waveguide entrance. Wavelength=10 um, and the input dimension of the Gaussian wave is 3 mm RMS.

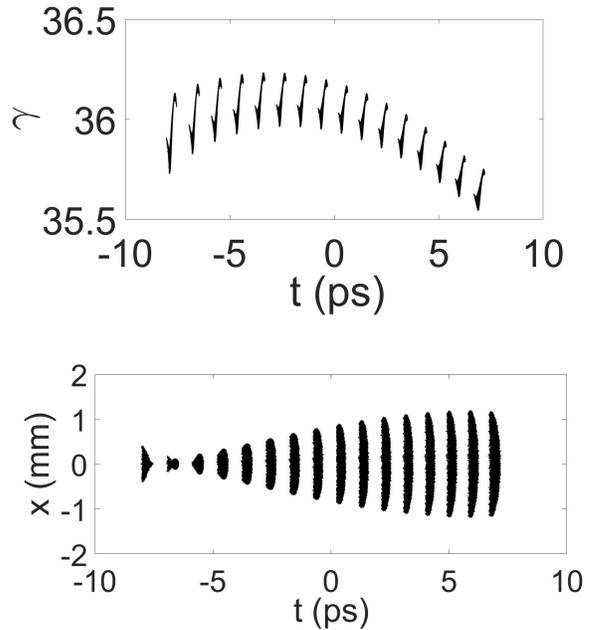

Fig.3 The longitudinal and transverse phase space of 16 microbunches at the entrance of the undulator

The undulator period length is 5.8 cm, the K value is 4.975 (corresponding to the fundamental resonant radiation frequency of 1THz ), the transverse size of the waveguide is 18×35 mm, which is made of aluminium whose electric conductivity is $3.816\times10^7$ S/m. Using the method presented in Sec II, we can obtain the time structure and spectrum in the fundamental mode of the output pulses at the undulator exit as shown in Fig. 4 and the transverse distribution of the total radiation power as shown in Fig.5.

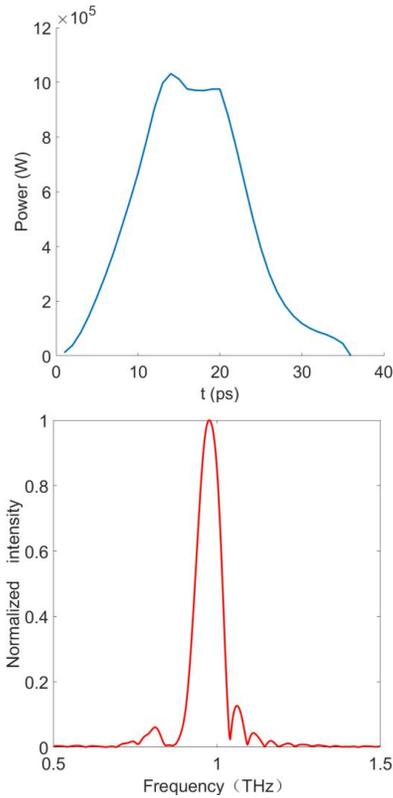

Fig. 4 Time structure and spectrum of the output field at the undulator exit

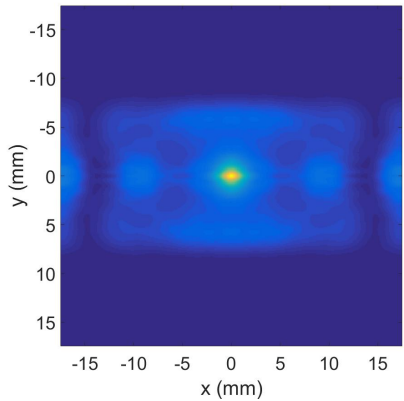

Fig. 5 Transverse distribution of the total radiation power

### IV. Summary and Discussion

In this paper, we propose a simple numerical method for the time-dependent, three-dimensional simulation of the free electron laser (FEL) using a rectangular waveguide within overmoded configuration. Although it is a special case, this condition can be satisfied for many practical far-infrared (FIR) and terahertz FEL facilities. This model also has the benefits of a small amount of modification of the GENESIS simulation code, which already has rich functions for FEL simulation and a wide range of users. There is no other code to check this method at present, however in terms of electromagnetic field propagation in a waveguide within overmoded configuration, good agreement with the numerical method using the fast Fourier transform can be obtained. Additionally, the format of the output data keeps the same with GENESIS, thus a combination of the modified GENESIS code for the FEL interaction and the OPC code [22] to model the FEL oscillators can still be used.

### V. ACKNOWLEDGMENTS

This work is supported by National Foundation of Natural Sciences of China (11705198, 11775216, 51627901, 11675178, 11611140102), China Postdoctoral Science Foundation (2017M622023), and Fundamental Research Funds for the Central Universities (WK2310000061).